\documentclass[reprint, amsmath,amssymb, aps]{revtex4-2}

\usepackage{graphicx}% Include figure files
\usepackage{dcolumn}% Align table columns on decimal point
\usepackage{bm}% bold math
\usepackage{epsfig}
\usepackage{color}
\usepackage{bbm}
\usepackage{longtable}
\usepackage{graphicx}% Include figure files
\usepackage{dcolumn}% Align table columns on decimal point
\usepackage{bm}% bold math
\usepackage{ulem}% strikethrough

\def\ketm#1{  \left\vert  #1   \right\rangle   }

\def\bram#1{  \left\langle  #1   \right\vert   }

\def\sprm#1#2{  \left\langle #1 \left\vert \right. #2 \right\rangle   }

\def\mem#1#2#3{  \left\langle #1 \left\vert  #2 \right\vert #3 \right\rangle   }

\usepackage[cp866]{inputenc}
\usepackage[english]{babel}
\usepackage{latexsym}
\usepackage{indentfirst}
\usepackage{amsmath}
\usepackage{amssymb}
 \usepackage{epsfig}
 \usepackage{graphicx}
 \usepackage{bm}

\newcommand{\be}{\begin{equation}}
 \newcommand{\ee}{\end{equation}}
\newcommand{\bear}{\be\begin{array}}
\newcommand{\bea}{\begin{eqnarray}}
\newcommand{\eea}{\end{eqnarray}}
\newcommand{\nn}{\nonumber}

\newcommand{\br}{{\bm r}}

\newcommand{\bk}{{\bm k}}

\usepackage[dvipsnames]{xcolor}

\newcommand{\dst}{\displaystyle}
\newcommand{\fr}[2]{\frac{{\dst #1}}{{\dst #2}}}

\begin{document}

\title{Wave function of a photon produced in the resonant scattering\\[0.2cm] of twisted light
by relativistic ions}

%
% ---------------------------- Authors -------------------------------------
%
\author{Dmitry~V.~Karlovets}
\affiliation{School of Physics and Engineering, ITMO University, RUS--199034, Saint-Petersburg, Russia}
% \affiliation{Tomsk State University, 634050 Tomsk, Russia}

\author{Valeriy~G.~Serbo}
\affiliation{Novosibirsk State University, RUS--630090, Novosibirsk, Russia}
\affiliation{Sobolev Institute of Mathematics, RUS--630090, Novosibirsk, Russia}

\author{Andrey Surzhykov}
\affiliation{Physikalisch--Technische Bundesanstalt, D--38116 Braunschweig, Germany}
\affiliation{Institut f\"ur Mathematische Physik, Technische Universit\"at Braunschweig, D--38106 Braunschweig, Germany}
\affiliation{Laboratory for Emerging Nanometrology Braunschweig, D-38106 Braunschweig, Germany}

\begin{abstract}
We present a theoretical investigation of the resonant elastic scattering of twisted light, carrying angular momentum, by partially stripped ions. Special emphasis is placed on a question of whether the scattered radiation is also twisted. In order to investigate such an ``angular momentum transfer'', we develop an approach that allows us to find a quantum state of the final photon without projecting it onto a detector state. A general expression for this so--called \textit{evolved} state of outgoing radiation is derived and it can be used to analyze the resonant scattering by any ion, independently of its shell structure. Here, we illustrate our approach with the strong electric dipole $n S_{0} \to n' P_{1} \to n S_{0}$ transitions, which play an important role for the Gamma Factory project at CERN. For the incident Bessel light, the scattered radiation is shown to be in a superposition of twisted modes with the projections of the total angular momentum $m_f = 0, \pm 1$. The larger values of $m_f$ can be efficiently generated by inducing transitions of higher multiplicity. This angular momentum transfer, together with a remarkable cross section that is many orders of magnitude larger than that of the backward Compton scattering, makes the resonant photon scattering an effective tool for the production of twisted x-- and even gamma--rays at the Gamma Factory facility.
\end{abstract}

\maketitle

\section{Introduction}

The resonant elastic photon scattering by an atom or an ion, which proceeds via formation and subsequent decay of a particular atomic state, has been in the focus of experimental and theoretical studies for many decades \cite{BeL82,KaK86,MaM00,LaS02,SaV20}. Recently, this process has attracted a particular attention also as a \textit{key element} for the Gamma Factory project at the CERN facility \cite{JaL18,BuC20,BuB21}. In the Gamma Factory, the incident laser radiation with the frequency $ \omega^{\rm (lab)}_i$ will be collided head--to--head with fast--moving partially stripped ions and scattered predominantly backwards. Owing to the Lorentz transformation between the laboratory (collider) and the ion--rest reference frames, the frequency of the scattered photons will be boosted to 
\be
\omega^{\rm (lab)}_f \approx 4 \gamma^2 \omega^{\rm (lab)}_i,
\ee
where $\gamma$ is the ion's Lorentz factor. Since ions with $\gamma =$~10--2900 can be stored in the LHC collider ring, the gamma--rays with the energies up to 30 MeV and the intensity up to $10^{17}$ photons per second can be generated from the incident laser radiation with an energy just about 1~eV, see Ref.~\cite{BuC20,BuB21} for further details.  

Besides energy and emission pattern, further characteristics of the resonantly scattered photons are also of great interest for the Gamma Factory. For example, a number of theoretical works have been performed recently to investigate the polarization of outgoing radiation \cite{SeS21,VoS21}. Yet another degree of freedom that also attracts particular attention is the orbital angular momentum (OAM) of light. The generation of photon beams, carrying a non--zero OAM projection onto their propagation direction, is presently under discussion in the framework of the Gamma Factory project. Having an energy of up to tens of MeV, these so--called \textit{twisted} beams can become a valuable tool for a wide range of studies in atomic, optical and nuclear physics \cite{Afa18,KnS18,Her17,Ros19}. 

The generation of twisted x-- or even gamma--rays at the Gamma Factory may rely on the resonant scattering of incident twisted light by ultra--relativistic partially stripped ions. However, in order to decide about the feasibility of this approach one has to investigate first how (and whether) the projection of the angular momentum is \textit{transferred} between incident and outgoing photons in the scattering process. Until now, the theoretical analysis of such a ``twistedness transfer'' has been based on \textit{projecting} the scattered light states onto the plane--, cylindrical-- (twisted), or spherical--wave solutions \cite{SeS21,TaN21}. The outcome of this projection can be understood as a probability to get a ``click'' in a \textit{detector} that is sensitive only to a particular state of light. In the present work, we propose an alternative and more general approach that allows one to perform a \textit{quantum tomography} of the scattered radiation without a need to introduce detector states. This approach makes use of the well--established S--matrix theory and provides the wave--function of the outgoing photon as it results from the scattering process itself.    

In order to discuss how the ``quantum tomography'' method helps to analyze the state of a photon, produced in the resonant scattering of incident twisted radiation, we have to remind first how to describe such (twisted) solutions. In Sec.~\ref{sec:plane_wave_twisted}, therefore, we briefly review the Bessel electromagnetic waves, which are characterized by a well--defined projection of their total angular momentum (TAM) onto a propagation direction and can be written as a coherent superposition of the standard plane--waves. The most general description of scattering of the Bessel photons is given in Sec.~\ref{sec:s-matrix} within the framework of the S--matrix theory. In particular, it is shown that the \textit{evolved} state of the outgoing light, obtained irrespectively of the detector setup, can be written as a sum of the plane--waves, weighted with the scattering matrix elements. In Sec.~\ref{sec:resonant_E1_scattering}, we evaluate these matrix elements and, hence, the vector potential of the final (evolved) photon state for the electric dipole $n S_{0} \to n' P_{1} \to n S_{0}$ resonant transition, which is of special interest for the Gamma Factory program. Based on the performed analysis, we argue that the resonant E1 scattering of the incident twisted light leads to the emission of secondary radiation in a superposition of Bessel states with the TAM projections $m_f = 0, \pm 1$. The larger values of $m_f$ can be obtained by inducing quadrupole and even higher--order ionic transitions. Moreover, as shown in Sec.~\ref{sec:compton}, the cross section of the resonant scattering is many orders of magnitude larger than that of the Compton scattering by free electrons; the process which is discussed as an alternative source of the high--energy twisted radiation \cite{JeS11,JeS11b}. Together with our predictions for the TAM transfer, this cross--section argument suggests that the resonant scattering by fast--moving partially stripped ions provides an efficient tool for generating x-- and even gamma--ray twisted photon beams at the Gamma Factory facility. The summary of these results is given finally in Sec.~\ref{sec:summary}.

Relativistic units ($\hbar =1,\; c = 1$) are used throughout the paper.

\section{Plane--wave and twisted states}
\label{sec:plane_wave_twisted}

Before starting theoretical investigation of the resonant scattering process, we have to remind first how to describe the \textit{states} of incident and outgoing photons. In this section we focus on two such states: (i) the ``standard'' plane--wave and (ii) the cylindrical--wave Bessel solutions. Since both the solutions and their relations are well discussed in the literature \cite{JeS11,JeS11b,MaH13,KnS18}, here we restrict ourselves just to a brief compilation of basic formulas, needed for the further analysis.   

\subsection{Plane--wave photons}
\label{subsec:plane_wave_photons}

The vector potential of a freely propagating electromagnetic plane--wave reads as:
\begin{equation}
    \label{eq:plane_wave_vector_potential}
    {\bm A}_{{\bm k} \lambda}({\bm r}) = 
    {\bm \epsilon}_{{\bm k} \lambda}\, e^{i {\bm k}\cdot{\bm r}} \, ,
\end{equation}
where ${\bm k}$ is the wave--vector, which defines not only the direction of propagation, $\hat{\bm k} = {\bm k}/k = \left(\theta, \varphi\right)$, but also the energy $\omega=|\bk|$ of light. In Eq.~(\ref{eq:plane_wave_vector_potential}), moreover, ${\bm \epsilon}_{{\bm k} \lambda}$ is the \textit{circular} polarization vector with $\lambda = \pm 1$ being the photon's helicity, i.e. projection of its spin onto the $\hat{\bm k}$--direction. Within the Coulomb gauge for the light--matter coupling, this polarization vector is orthogonal to the wave--vector, ${\bm k} \cdot {\bm \epsilon}_{{\bm k} \lambda} = 0$, and it can be written as:
\begin{equation}
    \label{eq:polarization_vector_expansion}
    \bm \epsilon_{\bk \lambda} = \sum\limits_{\sigma=0,\pm 1} e^{-i\sigma \varphi} \, d^{\;\;1}_{\sigma \lambda}(\theta) \, {\bm e}_{\sigma}.
\end{equation}
Here, $d^{\;\;1}_{\sigma \lambda}(\theta)$ are the small Wigner functions:
\begin{subequations}
\begin{eqnarray}
    d_{\lambda \lambda'}^{\;\,1}(\theta) &=& \fr 12 \left(1+\lambda \lambda'\cos\theta\right) \, , \\[0.2cm]
    d_{\lambda 0}^{\;1}(\theta) &=& -d_{0 \lambda}^{\;1}(\theta)=-\fr{\lambda}{\sqrt{2}} \sin\theta \, , \\[0.2cm]
    d_{00}^{\;1}(\theta) &=& \cos\theta \, ,
\end{eqnarray}
\end{subequations}
and the basis unit vectors:
\begin{equation}
    \label{eq:unit_vectors}
    {\bm e}_{0} = (0,0,1), \; \; {\rm and} \; \;
    {\bm e}_{\pm 1} = \frac{\mp 1}{\sqrt{2}} \, (1, \pm i, 0)\
\end{equation}
are the eigenvectors of the operator ${\hat S}_z$ of the spin projection onto $z$--axis:
\begin{equation}
    \label{eq:unit_vectors_eigenproblem}
    {\hat S}_z \, {\bm e}_{\sigma} = \sigma \, {\bm e}_{\sigma} \, ,
\end{equation}
with the eigenvalues $\sigma = 0, \pm 1$.

\subsection{Twisted Bessel photons}
\label{subsec:twisted_photons}

Besides the plane waves (\ref{eq:plane_wave_vector_potential}), we also briefly discuss the cylindrical--wave solutions of the free Maxwell equations as can be described, for example, by the so--called Bessel state. The Bessel beam, propagating along the quantization $z$--axis is characterized by its longitudinal momentum $k_z > 0$, an absolute value of the transverse momentum $k_\perp$, as well as by an energy $\omega=\sqrt{k^2_\perp+k^2_z}$. The vector potential of this Bessel photon state can be written as:
\begin{eqnarray}
    \label{eq:Bessel_state_along_z}
    {\bm A}_{k_\perp m k_z \lambda}(\br) &=& i^{-m}\,\int_0^{2\pi} \bm \epsilon_{\bk \lambda} \, e^{i\bk\cdot\br+im\varphi} \, \fr{d\varphi}{2\pi} \, ,
\end{eqnarray}
as a coherent superposition of plane waves with the helicity $\lambda$ and the wave--vectors ${\bm k} = \left(k_\perp \cos\varphi, k_\perp \sin\varphi, k_z \right)$. This expression implies, moreover, that the Bessel photon beam carries a well--defined $z$--projection
\begin{equation}
    m = 0, \pm 1, \pm 2, ...
\end{equation}
of its \textit{total} angular momentum (TAM). This can be easily obtained from Eq.~(\ref{eq:polarization_vector_expansion}) and the fact that $\bm \epsilon_{\bk \lambda}\, e^{im\varphi}$ is an eigenvector,
\begin{equation}
    \label{eq:Jz_projection}
    \hat J_z \, \bm \epsilon_{\bk \lambda}\, e^{im\varphi}= m\, \bm \epsilon_{\bk \lambda}\, e^{im\varphi}, \, 
\end{equation}
of the corresponding operator ${\hat J}_z$, which can be written as:
\begin{equation}
    \label{eq:Jz_definition}
    \hat J_z =-i\fr{\partial}{\partial \varphi} + \hat S_z \, ,
\end{equation}
where $\hat S_z \bm \epsilon_{\bk \lambda} = -i\,\bm \epsilon_{\bk \lambda} \times {\bm e}_{0}$, and ${\bm e}_{0}$ is given by Eq.(\ref{eq:unit_vectors}).

One can further evaluate the vector potential (\ref{eq:Bessel_state_along_z}) by performing integration over the azimuthal angle:
\begin{eqnarray}
    \label{eq:Bessel_state_along_z_2}
    {\bm A}_{k_\perp m k_z \lambda}(\br) &=& e^{ik_z z} \, \sum_{\sigma=0;\pm 1}  \Big[ i^{-\sigma} \, 
    d^{\;1}_{\sigma\,\lambda}(\theta) \nonumber \\[0.2cm]
    &\times& J_{m-\sigma}(k_\perp r_\perp) \, e^{i(m-\sigma)\varphi_r}\,
    {\bm e}_{\sigma} \Big] \, ,
\end{eqnarray}
where $J_{m-\sigma}(k_\perp r_\perp)$ is the Bessel function. As seen from this expression, the potential ${\bm A}_{k_\perp m k_z \lambda}(\br)$ depends on an opening angle:
\begin{equation}
    \label{eq:opening_angle_definition}
    \theta= \arctan(k_\perp/k_{z}) \, ,
\end{equation}
which is usually employed to characterize the ratio of the transverse to longitudinal momenta of twisted light. In the present--day experiments this ratio and, hence, the opening angle are usually rather small, $\theta \ll 1$, leading to the so--called \textit{paraxial} regime. In this regime, the sum in Eq.~(\ref{eq:Bessel_state_along_z_2}) is dominated by a single term with $\sigma=\lambda$ and the vector potential of the Bessel light can be written as:
\begin{eqnarray}
    \label{eq:Bessel_state_along_z_paraxial}
    {\bm A}_{k_\perp m k_z \lambda}(\br) &\approx& i^{-\lambda} \,\cos^2(\theta/2)\, J_{m-\lambda}(k_\perp r_\perp) \nonumber \\[0.2cm]
    &\times& {\rm e}^{i(m-\lambda)\varphi_r}\, {\bm e}_{\lambda} \, {\rm e}^{i k_z z} .
\end{eqnarray}
It implies for a paraxial twisted beam, that the $z$-projection of the total angular momentum $m$ unambiguously splits into the (projections of) spin $\lambda$, and the orbital $m-\lambda$ angular momenta. If, moreover, $m-\lambda\neq 0$, the energy density of the electromagnetic field vanishes in the center of such a beam, whereas the first maximum appears at the distance
$r_\perp \approx |m-\lambda|/(k \sin{\theta})$. 

In the limit of the vanishing transverse momentum and the opening angle, $k_\perp \to 0$ and $\theta \to 0$, the vector potential ${\bm A}_{k_\perp m k_z \lambda}(\br)$ can be further simplified to:
\begin{equation}
     {\bm A}_{k_\perp m k_z \lambda}(\br)\vert_{\theta \to\, 0} \approx \delta_{m\lambda} \,
 i^{-\lambda}\, {\bm e}_{\lambda} \, e^{i k z} \, ,
\end{equation}
where we used the well--known asymptotic formulas for the Wigner, $d^{\;1}_{\sigma \lambda}(0) = \delta_{\sigma \lambda}$, and the Bessel, $J_{m-\sigma}(0) = \delta_{m\,\sigma}$, functions. One can see from this expression that for $m = \lambda$ the vector potential of the Bessel photons restores, up to an insignificant phase factor $i^{-\lambda}$, the ``standard'' solution for a plane wave propagating along the $z$ axis.

Until now, we have discussed the basic expressions to describe the twisted Bessel photons in the \textit{coordinate} space. For the further analysis, however, it is useful to re--write the vector potential ${\bm A}_{k_\perp m k_z \lambda}$ also in the \textit{momentum} representation:
\begin{eqnarray}
    \label{eq:Bessel_state_along_z_momentum}
    \tilde{\bm A}_{k_\perp m k_z \lambda}({\bm p})&=&\int {\bm A}_{k_\perp m k_z \lambda}(\br)\, e^{-i{\bm p}\cdot\br} {\rm d}^3r \nonumber \\[0.2cm]
    &= & i^{-m} \, e^{im\varphi_p} \, \fr{4\pi^2}{p_\perp} \, {\bm \epsilon}_{{\bm p} \lambda} \nonumber \\
    &\times& \delta(p_\perp-k_\perp)\,
    \delta(p_z-k_z) .
\end{eqnarray}
This expression again suggests a ``visualization'' of the Bessel state as a superposition of plane waves, the wave--vectors of which lay on the surface of a cone with the opening angle $\theta$, see also Eq.~(\ref{eq:Bessel_state_along_z}).

\section{S--matrix approach}
\label{sec:s-matrix}

\subsection{General formalism}
\label{subsec:s-matrix-general}

Our further analysis is based on the standard S--matrix approach, which allows one to relate the states of a physical system long before and after the scattering. In the operator form, this relation can be written as follows:
\begin{equation}
    \label{eq:S-matrix-general}
    \ketm{e} = {\hat S} \ketm{i} \, ,
\end{equation}
where the explicit form of \textit{the evolution operator} ${\hat S}$ depends, of course, on a particular process. By defining such a process and by choosing the initial state $\ketm{i}$ of a system, one can use Eq.~(\ref{eq:S-matrix-general}) to \textit{obtain} the so--called evolved state $\ketm{e}$. 

It is important to distinguish \textit{a detected state} of the final photon $|f\rangle$, which is defined by the properties of a detector, and \textit{the evolved state} $|e\rangle$ in which the photon goes as a result of the scattering, irrespectively of the post-selection process. In this paper, we concentrate on the properties of the final photon itself, without a reference to how this photon is detected, which allows us to explicitly retrieve the phase of the evolved state and show that the latter might have a phase vortex. In other words, we perform \textit{quantum tomography} of the final photon state, while in order to judge whether the final photon is twisted or not it is usually enough to employ just two representations, the momentum and the coordinate one, corresponding to two marginal values of the quadrature operator $\hat{X}_{\alpha}$,
\begin{equation}
    \hat{X}_{0} = \hat{r}\ \text{and}\ \hat{X}_{\pi/2} = \hat{k}.
\end{equation}
Other intermediate values of the quadrature angle $\alpha$ are not needed for our purposes.

It is usually practical to re--write Eq.~(\ref{eq:S-matrix-general}) in terms of wave--functions, instead of state vectors. In order to achieve this one needs to agree first about the \textit{representation} in which the wave--functions are written. For example, the initial-- and final--state wave--functions of a system in the coordinate representation are related to each other as: 
\begin{eqnarray}
    \label{eq:S-matrix-general-wave-functions}
    \psi_{e}\left({\bm r}\right) &\equiv& \sprm{{\bm r}}{e} = \mem{{\bm r}}{{\hat S}}{i} = \sum\limits_{f} \sprm{{\bm r}}{f} \, \mem{f}{{\hat S}}{i} \nonumber \\[0.2cm]
    &\equiv&
    \sum\limits_{f} \psi_{f}\left({\bm r}\right) \, S_{fi} \, .
\end{eqnarray}
Here, we introduced the notation $S_{fi} = \mem{f}{{\hat S}}{i}$ for the S--matrix element and assumed that the detected states $\ketm{f}$ form a complete basis set, so that $\sum_f \ketm{f}\bram{f} = {\hat I}$ is the identity operator.

\subsection{Evolved photonic states}
\label{subsec:s-matrix-photons}

Having briefly discussed the general formulas of the S--matrix approach, we are ready to apply them to the photon scattering process. As seen from Eqs.~(\ref{eq:S-matrix-general}) and (\ref{eq:S-matrix-general-wave-functions}), this would require a definition of an initial photon state. By assuming, for example, the incident plane--wave with the wave--vector ${\bm k}_i$ and the helicity $\lambda_i$, one can derive
\begin{eqnarray}
    \label{eq:final_potential_plane_wave}
    {\bm A}_{e}^{\rm (pl)}({\bm r}) &=& \sum\limits_{f} \sprm{{\bm r}}{f} \, S^{\rm (pl)}_{f i}
    \nonumber \\[0.2cm]
    &=& \sum\limits_{\lambda_f{ = \pm 1}}  \int {\bm \epsilon}_{\bk_f \lambda_f} \, {\rm e}^{i\bk_f\cdot\br} \, S^{\rm (pl)}_{f i}\, \fr{{\rm d}^3 {\bm k}_f}{(2\pi)^3},
\end{eqnarray}
i.e., the vector potential of the evolved state of a scattered photon in the coordinate representation. The second line of this expression is obtained in the lowest order of the perturbation theory and by using complete basis set of the plane--wave solutions, $\sprm{{\bm r}}{f} = {\bm \epsilon}_{\bk_f \lambda_f} \, {\rm e}^{i\bk_f\cdot\br}$. In Eq.~(\ref{eq:final_potential_plane_wave}), moreover, the plane--wave scattering matrix element is given by:
\begin{eqnarray}
    \label{eq:s-matrix-element-plane-wave}
    S^{\rm (pl)}_{f i} &=& \mem{\bk_f \lambda_f}{{\hat S}}{\bk_i \lambda_i} \nonumber \\[0.2cm]
    &=& i \, N \, 8\pi^2 \, \delta(\omega_i - \omega_f) \, {\cal M}^{\rm (pl)}_{f i} \, ,
\end{eqnarray}
where the normalization factor $N = 1 /\sqrt{\left(2 \omega_i {\cal V}\right) \left(2 \omega_f {\cal V}\right)}$ corresponds to a single photon in a volume ${\cal V}$, and ${\cal M}^{\rm (pl)}_{f i}$ is a transition amplitude. This amplitude depends, of course, on a specific resonant transition in a ``target'' ion and is directly related to the differential cross--section of the resonant scattering:
\begin{equation}
    \label{eq:diff_cross_section}
    \fr{{\rm d}\sigma^{\rm (pl)}}{d\Omega_f}=\left|{\cal M}^{\rm (pl)}_{f i} \right|^2.
\end{equation}
While the explicit form of the plane--wave scattering amplitude ${\cal M}^{\rm (pl)}_{f i}$ will be discussed in the next Section, here we discuss scattering of the incident Bessel light.

Owing to the complex internal structure of twisted light \cite{KnS18,MaH13}, its scattering demands more attention, compared to the plane--wave case. One has to agree first about the \textit{geometry} of the scattering process. In the present work, we will consider a scenario of the head--to--head collisions of the incident Bessel and ion beams, which is planned to be realized at the Gamma Factory \cite{BuC20}. In this scenario, the twisted light \textit{counter--propagates} the quantization $z$--axis, chosen along the ion beam direction, and is described by the following vector potential:
\begin{eqnarray}
    \label{eq:Bessel_state_opposite_z}
    {\bm A}_{k_\perp m k_z \lambda}(\br) &=& 
    i^{m}\,\int_0^{2\pi} \bm \epsilon_{\bk \lambda}\, e^{i\bk\cdot\br-im\varphi}\fr{d\varphi}{2\pi} \nonumber \\[0.2cm]
    && \hspace*{-1cm} = (-1)^m\,e^{ik_z z} \sum_{\sigma=0;\pm 1} \Big[ i^{\sigma} d^{\;1}_{\sigma\,\lambda}(\theta) \nonumber \\
    && \hspace*{-1cm} \times J_{m+\sigma}(k_\perp r_\perp)\,e^{-i(m+\sigma)\varphi_r}\,
    {\bm e_{\sigma}} \Big] \, ,
\end{eqnarray}
where the longitudinal momentum is negative, $k_z < 0$, and the opening angle lies in the interval $\pi/2 < \theta < \pi$. One can note that this expression is trivially obtained from Eqs.~(\ref{eq:Bessel_state_along_z}) and (\ref{eq:Bessel_state_along_z_2}) by changing the sign of $m$. 

With the help of Eq.~(\ref{eq:Bessel_state_opposite_z}) one can derive the vector potential of the scattered photon for the incident Bessel light. Indeed, by applying the S--matrix approach we easily find it in the coordinate representation: 
\begin{eqnarray}
    \label{eq:final_potential_twisted_wave}
    {\bm A}_{e}^{\rm (tw)}({\bm r}) &=&  \sum\limits_{f} \sprm{{\bm r}}{f} \, S^{\rm (tw)}_{f i} \nonumber \\[0.2cm]
    &=& \sum\limits_{\lambda_f{ = \pm 1}}  \int {\bm \epsilon}_{\bk_f \lambda_f} \, {\rm e}^{i\bk_f\cdot\br} \, S^{\rm (tw)}_{f i}\, \fr{{\rm d}^3 {\bm k}_f}{(2\pi)^3} \, ,
\end{eqnarray}
where we again perform ``summation'' over the complete set of the plane--wave solutions $\sprm{{\bm r}}{f} = {\bm \epsilon}_{\bk_f \lambda_f} \, {\rm e}^{i\bk_f\cdot\br}$ in the first line, and use the lowest order perturbation theory in the second line. Moreover, the vector potential in the momentum representation
\begin{eqnarray}
    \label{eq:final_potential_twisted_wave_momentum}
    \tilde{\bm A}^{\rm (tw)}_e({\bm k}_e) &=& \int {\bm A}_{e}^{\rm (tw)}({\bm r}) \, {\rm e}^{-i {\bm p}\cdot\br} \, {\rm d}^3 {\bm r} \nonumber \\[0.2cm]
    &=& \sum_{\lambda_f{ = \pm 1}}  {\bm \epsilon_{\bk_f \lambda_f}} \, S^{\rm (tw)}_{f i} \, \Big|_{{\bm k}_f = {\bm k}_e} \, , 
\end{eqnarray}
immediately follows from Eq.~(\ref{eq:final_potential_twisted_wave}), and has even simpler form.

As seen from Eqs.~(\ref{eq:final_potential_twisted_wave}) and (\ref{eq:final_potential_twisted_wave_momentum}), the analysis of the scattering of initially Bessel radiation is traced back to the S--matrix element
\begin{eqnarray}
    \label{eq:s-matrix-element-twisted-1}
    S^{\rm (tw)}_{f i} &=& \mem{\bk_f \lambda_f}{{\hat S}}{k_{\perp, i} \, m_i \, k_{z, i} \, \lambda_i} \, ,
\end{eqnarray}
which describes ``transition'' between the twisted $\ketm{k_{\perp, i} \, m_i \, k_{z, i} \, \lambda_i}$ and the plane--wave $\ketm{\bk_f \lambda_f}$ photon states. This matrix element can be again expressed in terms of the following transition amplitude:
\begin{equation}
    \label{eq:s-matrix-element-twisted-2}
    S^{\rm (tw)}_{f i} = i \, N^{\rm (tw)} \,  8\pi^2 \delta(\omega_i - \omega_f) \,
    {\cal M}^{\rm (tw)}_{f i}({\bm b}) \, \, ,
\end{equation}
which, in turn, is directly related to its plane--wave counterpart,
\begin{eqnarray}
    \label{eq:s-matrix-element-twisted-3}
    {\cal M}^{\rm (tw)}_{f i}({\bm b}) && \nonumber \\[0.2cm]
    && \hspace*{-1.2cm} = \int_0^{2\pi} \fr{{\rm d}\varphi_i}{2\pi} \,
    i^{m_i} e^{-i m_i \varphi_i +i {\bm b}\cdot\bk_{\perp, i}} 
    {\cal M}_{fi}^{\rm (pl)}(\varphi_i) \, .
\end{eqnarray}
Here, we have introduced \textit{an impact parameter} 
$$
{\bm b} = b (\cos\varphi_b,\, \sin\varphi_b,\,0)
$$
to describe the displacement of the axis of the incident Bessel beam from a ``target'' ion. In Eq.~(\ref{eq:s-matrix-element-twisted-2}), moreover, the normalization factor $N^{\rm (tw)}=\sqrt{k_\perp / 4\omega_i {\cal RL}_z \, 2\omega_f \cal V}$ corresponds to a single photon in the cylindrical volume ${\cal V}=\pi{\cal R}^2 {\cal L}_z$.

We conclude from the analysis above that in order to retrieve the final photon state, both for the incident plane--wave and twisted radiation, the knowledge of the (plane--wave) amplitude ${\cal M}_{fi}^{\rm (pl)}$ is required. The explicit form of this amplitude depends, of course, on details of the electronic structure of a ``target'' ion as well as on a particular resonant transition. In the next Section, for example, we will discuss the electric dipole ($E1$) scattering channel $n S_{0} \to n' P_{1} \to n S_{0}$, for which the amplitude ${\cal M}_{fi}^{\rm (pl)}$ has a simple analytic form.

\section{Resonant $E1$ scattering}
\label{sec:resonant_E1_scattering}

In order to illustrate the application of the general expressions (\ref{eq:final_potential_plane_wave}) and (\ref{eq:final_potential_twisted_wave})--(\ref{eq:final_potential_twisted_wave_momentum}) for the vector potentials of scattered photons, we focus here on the photon scattering off an ion, being initially in a positive--parity state with the total angular momentum $J_i = 0$, and which proceeds via intermediate ionic state with $J_\nu^{P_\nu}
= 1^-$. The examples of this $E1$ scattering channel are the $1s \to 2p \to 1s$ transition in non--relativistic one--electron ions and $n S_{0} \to n' P_{1} \to n S_{0}$ transitions in various many--electron systems. We will assume, moreover, the scattering geometry as will be observed in the Gamma Factory experiments \cite{BuC20}, where the incident photon beam counter--propagates the ion beam direction, and where the latter is chosen as quantization $z$--axis.

\subsection{Scattering of plane-wave photons}
\label{subsec:plane_wave_scattering}

We start our analysis from the well--known case of the $n S_{0} \to n' P_{1} \to n S_{0}$ resonant scattering of incident plane--wave light. The amplitude for this transition has been discussed in the literature, see~\cite{MaM00,SeS21}, and can be written as:
\begin{equation}
    \label{eq:amplitude_E1_plane}
    {\cal M}^{\rm (pl)}_{f i} = R(\omega_i) \, 
    ({\bm \epsilon}^*_{\bk_f \lambda_f} \cdot {\bm \epsilon}_{\bk_i \lambda_i}),
\end{equation}
where ${\bm \epsilon}_{\bk_i \lambda_i}$ and ${\bm \epsilon}_{\bk_f \lambda_f}$ are polarization vectors of initial-- and final--state photons, and the radial function:
\begin{equation}
    \label{eq:Rnu_function}
    R(\omega_i) = \fr{3}{2\omega_i}\,\fr{\Gamma_\nu/2}{E_{\nu i}-\omega_i - i\Gamma_\nu/2} \, 
\end{equation}
that depends only on the energies of the incident photon $\omega_i$ and of the transition $E_{\nu i} = E_{\nu} - E_i$ as well as on the width of the excited ionic state.

By inserting the amplitude (\ref{eq:amplitude_E1_plane}) into Eqs.~(\ref{eq:final_potential_plane_wave}) and (\ref{eq:s-matrix-element-plane-wave}) we obtain the vector potential of evolved (scattered) photons in the coordinate representation:
\begin{eqnarray}
    \label{eq:final_potential_plane_wave_final_1}
    {\bm A}_{e}^{\rm (pl)}({\bm r}) &=& i \, N \, 8\pi^2 \, R(\omega_i) \nonumber \\[0.2cm]
    && \hspace*{-1.5cm} \times \int {\bm \chi}_{-\lambda_i}({\bm k}_f) \, {\rm e}^{i\bk_f\cdot\br} \, \delta\left(\omega_i - \omega_e \right)\fr{{\rm d}^3 {\bm k}_f}{(2\pi)^3} \, .
\end{eqnarray}
Here we used the fact that for the head--on collisions $\bm \epsilon_{\bk_i \lambda_i} = {\bm e}_{-\lambda_i}$, as follows from Eq.~(\ref{eq:polarization_vector_expansion}) for $\theta_i = \pi$ and $\varphi_i = 0$, and introduced the vector:
\begin{eqnarray}
    \label{eq:chi_vector}
    {\bm \chi}_{\sigma}({\bm k}_f) &=& \sum\limits_{\lambda_f} {\bm \epsilon}_{\bk_f \lambda_f} \, \left({\bm \epsilon}^*_{\bk_f \lambda_f} \cdot {\bm \epsilon}_{\bk_i -\sigma}\left(\theta_i = \pi \right)\right)
    \nonumber \\[0.2cm]
    &=& \sum\limits_{\lambda_f} {\bm \epsilon}_{\bk_f \lambda_f} \, \left({\bm \epsilon}^*_{\bk_f \lambda_f} \cdot {\bm e}_{\sigma} \right)
    \nonumber \\[0.2cm]
    &=& {\rm e}^{i \sigma\varphi_f } \, \sum\limits_{\lambda_f} {\bm \epsilon}_{\bk_f \lambda_f} \, d^1_{\sigma \lambda_f}(\theta_f) \, ,
\end{eqnarray}
with $\sigma = 0, \, \pm 1$ and  ${\lambda_f = \pm 1}$.

By making use of Eq.~(\ref{eq:final_potential_plane_wave_final_1}) one can easily derive the vector potential in the momentum representation:
\begin{equation}
    \label{eq:final_potential_plane_wave_final_momentum}
     {\tilde {\bm A}}_{e}^{\rm (pl)}({\bm k}_e) = 2\pi N \, i \, \delta(\omega_i - \omega_e)
    R(\omega_i) \, {\bm \chi}_{-\lambda_i}({\bm k}_e) \, .
\end{equation}
Apart of its simple analytical form, this expression allows one to analyze the properties of the scattered photon. Indeed, by employing the definition (\ref{eq:chi_vector}) and Eqs.~(\ref{eq:Jz_projection})--(\ref{eq:Jz_definition}) we trivially obtain:
\begin{subequations}
\begin{eqnarray}
    && {\bm k}_e \cdot \tilde {\bm A}_{e}^{\rm (pl)}({\bm k}_e) = 0 \, , \\[0.2cm]
    && {\hat J}_z \tilde {\bm A}_{e}^{\rm (pl)}({\bm k}_e) = -\lambda {\tilde {\bm A}}_{e}^{\rm (pl)}({\bm k}_e) \, ,\\[0.2cm]
    && {\hat {\bm J}}^2 {\tilde {\bm A}}_{e}^{\rm (pl)}({\bm k}_e) = 2 {\tilde {\bm A}}_{e}^{\rm (pl)}({\bm k}_e) \, ,
\end{eqnarray}
\end{subequations}
which implies that the vector potential ${\tilde{\bm A}}_{e}^{\rm (pl)}({\bm k}_e)$ represents a \textit{spherical wave} with the energy $\omega_e = \omega_i$ and a definite value of the total angular momentum, $J_f = 1$. This \textit{dipole} wave possesses, moreover, a well--defined projection $J_z$ onto the quantization axis:
\begin{equation}
    m_e = -\lambda_i \, ,
\end{equation}
as can be expected from the angular momentum conservation rule.

Finally, by projecting the vector potential (\ref{eq:final_potential_plane_wave_final_momentum}) onto the plane--wave solution and using Eq.~(\ref{eq:chi_vector}), one can again restore the well--known angular distribution:
\begin{eqnarray}
    W(\theta_f, \varphi_f) &=& \left| \sprm{ {\tilde {\bm A}}_{{\bm k}_f \lambda_f}^{\rm (pl)}}{{\tilde {\bm A}}_{e}^{\rm (pl)}} \right|^2 \nonumber \\[0.2cm]
    &\propto& \left|({\bm \epsilon}^*_{\bk_f \lambda_f} \cdot {\bm \epsilon}_{\bk_i \lambda_i}) \right|^2 \, ,
\end{eqnarray}
of the photons, emitted in the $n S_{0} \to n' P_{1} \to n S_{0}$ resonant transition, see Ref.~\cite{SeS21}. 

\subsection{Scattering of twisted photons}
\label{subsec:twisted_wave_scattering}

As the second example, we investigate the vector potential of light, emitted in the $n S_{0} \to n' P_{1} \to n S_{0}$ resonant transition for the incident Bessel photons. As has already been mentioned, a special attention has to be paid for this case to the \textit{geometry} of the process and, in particular, to the choice of the quantization ($z$--) axis. Below we consider two particular choices of the $z$--axis, which can be employed for the theoretical description of {\it the same} evolved photon state. First, we study the resonant scattering for the geometry in which a ``target'' ion is located on the quantization $z$--axis, while the axis of the incident twisted beam is anti--parallel to $z$ and is shifted from it by the impact parameter $\bm b$. In the second scenario, we choose the $z$--axis to pass through the center of the Bessel beam in the direction opposite to its propagation, and displace instead the ion by $\bm b$. Of course, both the choices coincide for the vanishing impact parameter $b=0$, and this special case will be discussed in Sec.~\ref{subsubsec:b_0}. Finally, a scenario of collision of the \textit{finite--size} Bessel photon and ion beams is recalled briefly in the end of this Section. 

%\begin{widetext}
\begin{figure*}[t]
	%\hspace*{-0.2cm}
	\center
	\includegraphics[width=1.0\linewidth]{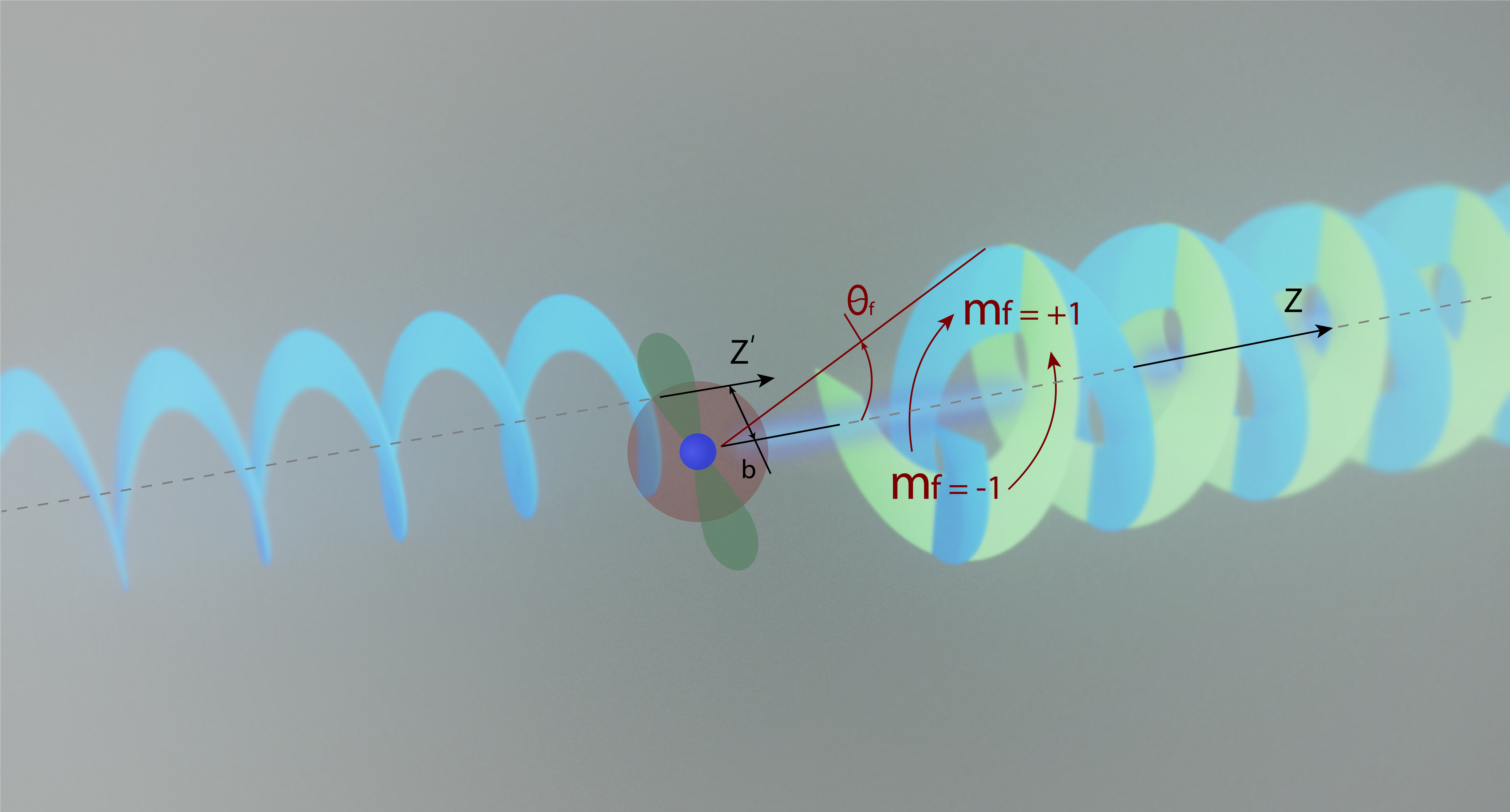}
	\caption{Resonant $E1$ scattering of a twisted photon by an atom or an ion, during which a transition $n S_{0} \to n' P_{1} \to n S_{0}$ (s- and p-orbitals are schematically shown) takes place. The ion's center is located on the z axis, while $b$ is an impact-parameter. The final photon itself is in a superposition of states with the total angular momenta $m_f =0, +1, -1$, according to Eq.(\ref{eq:final_potential_twisted_wave_final_coordinate}) or Eq.(\ref{eq:final_potential_twisted_wave_momentum_resonant}).}
\label{Fig_Sc}
\end{figure*}
%\end{widetext}

%
%
\subsubsection{Ion is located on the $z$--axis}
\label{subsubsect:ion_z_axis}

We start our analysis from the transition amplitude
\begin{eqnarray}
    \label{eq:scattering_amplitude_twsted}
    {\cal M}^{\rm (tw)}_{f i}({\bm b}) &=& R(\omega_i) \, \int_0^{2\pi} \, \fr{d\varphi_i}{2\pi} \, \Big\{
    i^{m_i} \, e^{-i m_i \varphi_i + i {\bm b}\cdot\bk_{\perp,i}} \nonumber \\[0.2cm] 
    && \hspace*{1cm} \times ({\bm \epsilon}^*_{\bk_f \lambda_f} \cdot {\bm \epsilon}_{\bk_i \lambda_i}) \, \Big\} \, ,
\end{eqnarray}
which is obtained from Eq.~(\ref{eq:s-matrix-element-twisted-3}) and the plane--wave counterpart (\ref{eq:amplitude_E1_plane}). By expanding the scalar product of the polarization vectors in terms of the small Wigner functions, 
\begin{eqnarray}
    ({\bm \epsilon}^*_{\bk_f \lambda_f} \cdot {\bm \epsilon}_{\bk_i \lambda_i}) &&
    \nonumber \\[0.2cm]
    && \hspace*{-2cm} = \sum_{\sigma=0;\pm 1} e^{i\sigma(\varphi_f - \varphi_i)}\,
    d^{\;1}_{\sigma \lambda_i}(\theta_i) \, d^{\;1}_{\sigma \lambda_f}(\theta_f) \, ,
\end{eqnarray}
and by performing integration over $\varphi_i$ we can further evaluate this amplitude as: 
\begin{eqnarray}
    \label{eq:scattering_amplitude_twsted_2}
    {\cal M}_{f i}^{\rm (tw)}({\bm b}) &=& R(\omega_i) \, (-1)^{m_i} \, 
    {\rm e}^{-i m_i \varphi_b} \, \sum_\sigma \Big[ 
    i^{\sigma} \, {\rm e}^{i\sigma(\varphi_f - \varphi_b)} \nonumber \\[0.2cm] 
    &\times& J_{m_i + \sigma}(k_{\perp, i} \, b) \, d^{\;1}_{\sigma \lambda_i}(\theta_i) \, d^{\;1}_{\sigma\lambda_f}(\theta_f) \Big] \, .
\end{eqnarray}
Here, $\varphi_f$ and $\varphi_b$ are azimuthal angles of the photon momentum ${\bk}_f$ and of the impact parameter vector ${\bm b}$, respectively.

With the help of the matrix element (\ref{eq:scattering_amplitude_twsted_2}) and Eq.~(\ref{eq:final_potential_twisted_wave}) we obtain, after simple algebra, the vector potential of the evolved photon state:
\begin{eqnarray}
    \label{eq:final_potential_twisted_wave_final_coordinate}
    {\bm A}_{e}^{\rm (tw)}({\bm r}; \, m_i) &=& (-1)^{m_i} \, \fr{i}{2\pi} \, \omega_i^2 \, N^{\rm (tw)} \, R(\omega_i) \, {\rm e}^{-i m_i \varphi_b} \nonumber \\[0.2cm]
    && \hspace*{-2.2cm} \times \sum_{\lambda_f = \pm 1}\sum_{m_f = 0,\pm 1} \int_0^\pi  C_{\lambda_f m_f}{({\bm b})} {\bm A}_{k_{\perp, f} \, m_f \, k_{z, f} \, \lambda_f}({\bm r})\nonumber \\[0.2cm]
    && \hspace*{-2.2cm}\times
    \,\sin\theta_f {\rm d}\theta_f \, , 
\end{eqnarray}
where the potential ${\bm A}_{k_{\perp, f} \, m_f \, k_{z, f} \, \lambda_f}({\bm r})$ is defined in Eq.~(\ref{eq:Bessel_state_along_z}) and the coefficient $C_{\lambda_f m_f}$ reads as follows:
\begin{eqnarray}
    \label{eq:C_coefficient}
     C_{\lambda_f m_f}{({\bm b})} &=& (-1)^{m_f} \, e^{-im_f \varphi_b} J_{m_i+m_f}(k_{\perp, i} \, b)
     \nonumber \\[0.2cm]
     &\times& d^{\;1}_{m_f \lambda_i}(\theta_i) \, d^{\;1}_{m_f \lambda_f}(\theta_f) \, .
\end{eqnarray}
Similar to the plane--wave case, the vector potential can also be derived in the momentum space:
\begin{eqnarray}
    \label{eq:final_potential_twisted_wave_momentum_resonant}
    \tilde{\bm A}^{\rm (tw)}_e({\bm k}_e; m_i) &=& 2 \pi i \, (-1)^{m_i} \, N^{\rm (tw)} \, 
    \delta (\omega_i - \omega_e) \, R(\omega_i) \nonumber \\[0.2cm]
    && \hspace*{-1.3cm} \times {\rm e}^{-i m_i \varphi_b} \, \sum\limits_{m_f = 0, \pm 1} \, 
    \Big[ i^{m_f} \, {\bm \chi}_{m_f}({\bm k}_e) \nonumber \\[0.2cm] 
    && \hspace*{-1.3cm} \times  d_{m_f \lambda_i}^{\;1}(\theta_i) \, J_{m_i + m_f}(k_{\perp, i} \, b) \, {\rm e}^{-i m_f \varphi_b} \Big] \, ,
\end{eqnarray}
with the vector ${\bm \chi}_{m_f}$ defined in Eq.~(\ref{eq:chi_vector})

Eqs.~(\ref{eq:final_potential_twisted_wave_final_coordinate}) and (\ref{eq:final_potential_twisted_wave_momentum_resonant}) allow one to analyze the properties of the scattered photons \textit{without} 
a need for projecting the evolved photon state onto any detector state. For example, we conclude that the outgoing radiation can be seen as a \textit{coherent superposition} of the Bessel states with the TAM projections: 
\begin{equation}
    \label{eq:mf_selection_rule}
    m_f = 0, \pm 1 \, 
\end{equation}
for any initial $m_i$. 
\subsubsection{Ion is displaced from the $z$--axis}

For the analysis of the scattering process under realistic experimental conditions it might be more convenient to choose the coordinate system in a different way than it is done in Subsection~\ref{subsubsect:ion_z_axis}. Namely, we can let the $z$--axis to pass through the center of the Bessel beam in the direction opposite to its propagation, and displace instead the ion by the impact parameter $\bm b$. For this choice of the quantization axis, one can still apply Eq.~\eqref{eq:final_potential_twisted_wave_final_coordinate}, in the right--hand side of which we have to expand the basis wave vectors ${\bm A}_{k_{\perp, f} \, m_f \, k_{z, f} \, \lambda_f}({\bm r})$ in terms of their ``on--axis'' counterparts
\begin{eqnarray}
    \label{eq:final_potential_twisted_wave_final_coordinate_shifted}
    &&{\bm A}_{k_{\perp, f} \, m_f \, k_{z, f} \, \lambda_f}({\bm r}) \nn
    \\
    &=& \sum_{m_f'=-\infty}^{\infty}
    \Big[ e^{-i(m_f' - m_f) \varphi_b} \, J_{m_f'- m_f}(k_{\perp, i} \, b) \nonumber \\[0.2cm]
    &\times& {\bm A}^{\rm(b)}_{k_{\perp, f} \, m'_f \, k_{z, f} \, \lambda_f}({\bm r}) \Big],
\end{eqnarray}
see Appendix C in Ref.~\cite{IvS16} for further details. We note that this expansion can be obtained by multiplying the vector potential in the momentum representation (\ref{eq:Bessel_state_along_z_momentum}) by a ``translation'' factor $\exp(i{\bm b}\cdot{\bm k}_e)$ and by expanding the exponent into series over the Bessel functions. It should be stressed that for the small values of $b < 1/k_{\perp, i}$ the distribution over $m'_f$ is concentrated near the values $m_f$.

\subsubsection{Zero impact parameter, $b=0$}
\label{subsubsec:b_0}

Of course, for $b = 0$ both the coordinate systems coincide and the evolved wave function is simplified. Indeed, in this case $J_{m_i + m_f}(0) = \delta_{m_i -m_f}$ and  
$J_{m'_f - m_f}(0) = \delta_{m_f m'_f}$, 
and the initial-- and final--state TAM projections are related to each other as
\begin{equation}
    \label{eq:mf_selection_rule_b0}
    m_f = - m_i \, ,
\end{equation}
as can be expected from the conservation rules. This relation, together with Eq.~(\ref{eq:mf_selection_rule}), suggests that only the incident Bessel photons with $m_i = 0, \pm 1$ can induce $n S_0 \to n' P_1$ (E1) excitations of an ion, placed at $b = 0$. Again, it fully meets our expectations based on the electric--dipole selection rules. 

Despite the fact that for $b = 0$ the scattered photon wave carries a well--defined projection of the total angular momentum onto the $z$--axis, see Eq.~(\ref{eq:mf_selection_rule_b0}), it can \textit{not} be described by a single pure Bessel state. This immediately follows from Eq.~(\ref{eq:final_potential_twisted_wave_final_coordinate}) which simplifies, for the vanishing impact parameter, to
\begin{eqnarray}
    \label{eq:final_potential_twisted_wave_final_coordinate_b0}
    {\bm A}_{e}^{\rm (tw)}({\bm r}; \, m_i) &=& \fr{i}{2\pi} \, \omega_i^2 \, N^{\rm (tw)} \, 
    R(\omega_i) \, d^{\;\;\;1}_{-m_i, \lambda_i}(\theta_i) \nonumber \\[0.2cm]
    && \hspace*{-0.5cm} \times \sum_{\lambda_f} \int_0^\pi {\rm d}\theta_f \,
    \Big\{ {\bm A}_{k_{\perp, f} \, -m_i \, k_{z, f} \, \lambda_f}(\br) \nonumber \\[0.2cm]
    && \hspace*{-0.5cm} \times d^{\;\;\;1}_{-m_i,\lambda_f}(\theta_f) \,\sin\theta_f 
    \Big\} \, ,
\end{eqnarray}
and suggests that outgoing photons acquire some momentum distribution. One may note that similar momentum distribution can be expected also for scattering of the Laguerre--Gaussian beams or the normalized wave--packets, similar to those discussed in Ref.~\cite{IvS11}. 

Beside the limit of the vanishing impact parameter $b$, we also briefly recall the case $\theta_i \to \pi$ and, hence, $k_{\perp, i} \to 0$. Most easily, this regime can be analyzed in the momentum space, in which
\begin{equation}
    d^{\;1}_{m_f \lambda_i}(\theta_i) \, J_{m_i + m_f}(k_{\perp,i} b) \approx \delta_{-m_i \lambda_i} \, \delta_{-m_i m_f} \, ,
\end{equation}
and the vector potential of the scattered photons (\ref{eq:final_potential_twisted_wave_momentum}) coincides with the standard plane-wave result (\ref{eq:final_potential_plane_wave_final_momentum}) up to an inessential factor $i^{\lambda_i}$.

\subsubsection{Collisions of beams of final width}

%As mentioned already above, the vorticity (``twistedness'') of incident Bessel photon is fully %transferred in this case to the final photon, c.f.~Eq.~(\ref{eq:mf_selection_rule_b0}). This can %be seen as a consequence of the axial symmetry of the entire system ``ion + light'' for $b =0$, %that results in the conservation of the $z$ projection of the total angular momentum. However, %this symmetry is \textit{broken} if an ion is displaced from the center of the photon beam, $b %\ne 0$, which leads to some distribution over the final--state projections $m'_f$. 

Until now, we have discussed the resonant scattering of twisted light by a well--localized \textit{single} atom. However, the derived expressions can be extended to analyze collisions of the Bessel photon and ion \textit{beams}, having finite cross sectional areas of $1/k_{\perp,i}$ and $\sigma_{\rm ion}$, respectively. Of special interest here is, again, the TAM projections of the outgoing photons. Owing to the finite (transverse) size of an ion beam, some of its ions will be displaced from the center of the Bessel beam, $b \ne 0$, thus resulting in a breakdown of relation (\ref{eq:mf_selection_rule_b0}) and, hence, in a distribution of the TAM values. Of course, the shape and width of this distribution depend on a particular experimental setup. If, for example, the transverse size of an ion beam $\sigma_{\rm ion}$ is smaller than \textit{the transverse coherence length} $1/k_{\perp, i}$ of the incident Bessel light, and the centres of the colliding beams (almost) coincide, the TAM distribution of scattered photons will be peaked around $m_f = - m_i$ with the following width:
\begin{eqnarray}
    \label{eq:width_TAM_distribution}
    \delta m_f &=& \sqrt{\langle\hat{J}_{z}^2\rangle_e - \langle\hat{J}_{z}\rangle^2_e} \nonumber \\[0.2cm]
    &\sim& \sigma_{\rm ion} \, k_{\perp, i} = \sigma_{\rm ion} \, k_{z, i} \, \tan\theta_i \, .
\end{eqnarray}
This estimate follows from the fact that the wave function (\ref{eq:final_potential_twisted_wave_final_coordinate}) or (\ref{eq:final_potential_twisted_wave_final_coordinate_shifted}) is proportional to $J_{m_i + m_f}(k_{\perp}b)$, where for a beam, treated as an \textit{incoherent} mixture of ions, the maximum value of the impact parameter is of the order of the beam width, $b \sim \sigma_{\rm ion}$. For the Gamma Factory project, if the width of the ion beam $\sigma_{\rm ion}$ is much smaller than the transverse coherence length of the incoming photons, the width $\delta m_f$ is vanishing. Say, for Pb ions it is $\sigma_{\rm ion} \approx 16\, \mu\text{m}$ \cite{BuC20} and a paraxial optical photon beam satisfies this condition.

\section{Compton scattering}
\label{sec:compton}

It follows from the analysis above that the resonant light scattering by fast--moving partially stripped ions may lead to the emission of high--energy twisted photon beams in the laboratory (collider) frame. Yet another possibility for the production of such beams was discussed recently in the literature, Refs.~\cite{JeS11, JeS11b}, and it relies on the inverse Compton scattering off free electrons: 
\begin{equation}
    \gamma({\bm k}_i) + e({\bm p}_i)\to \gamma({\bm k}_f)+ e({\bm p}_f) \, .
\end{equation}
If the energy of the incident photon is much smaller than the electron mass, the ``plane--wave'' matrix element of this process:
\begin{equation}
    \label{eq:Compton_scattering}
    {\cal M}^{\rm (C)}_{fi}= r_e \, (\bm \epsilon^*_{\bk_f \lambda_f} \cdot \bm \epsilon_{\bk_i \lambda_i}),
\end{equation}
has the same structure as the resonant--scattering one (\ref{eq:amplitude_E1_plane}), and where
$r_e$ is the classical electron radius. By making use of this matrix element one can investigate the properties of the Compton--scattered photons when incident light is prepared in the Bessel states, see Refs.~\cite{JeS11,JeS11b} for further details. It was shown, in particular, that almost complete ``TAM transfer'' between incoming and outgoing photons, $m_f \approx - m_i$, can be observed for small scattering angles,
typical for ultra--relativistic electrons. Moreover, in contrast to the resonant photon scattering, there is no selection rule similar to (\ref{eq:mf_selection_rule}) for the Compton case, which implies that high--energy photons with \textit{any} TAM projection can potentially be generated in the process (\ref{eq:Compton_scattering}). 

A great advantage of the resonant scattering by partially stripped ions, however, is that its cross section is about $6-12$ orders of magnitude \textit{larger} than that of the Compton scattering \cite{SeS21}. This can be easily seen from Eqs.~(\ref{eq:amplitude_E1_plane}) and (\ref{eq:Compton_scattering}), and from the fact that the cross sections of the scattered processes are determined by the squares of their transition amplitudes. Indeed, while the Compton cross section is proportional to the square of the classical electron radius, $\sigma^{\rm (C)} \propto r_e^2 \approx 7.8 \, \times 10^{-26}$~cm$^2$, its resonant scattering counterpart is $\sigma^{\rm (res)} \propto \left| R(\omega_i) \right|^2$, with the radial function $R(\omega_i)$ given by Eq.~(\ref{eq:Rnu_function}). For vanishing \textit{detuning}, $E_{\nu i} - \omega_i = 0$, this function is maximal and it gives:
\begin{eqnarray}
    \label{eq:Rnu_maximum}
    \left| R(\omega_i = E_{\nu i}) \right|^2 &=& \frac{9}{4 \omega^2_i} = \frac{9}{16 \pi^2}
    \lambda^2_{i} \nonumber \\[0.2cm] 
    && \hspace*{-2cm} \approx \left(8.6 \times 10^{-20} \div 1.6 \times 10^{-14}\right) \, {\rm cm}^2 \, ,
\end{eqnarray}
where $\lambda_i = 2\pi/\omega$ is the wave--length of the incident light and the values in the second line are obtained for the typical experimental scenarious of the future Gamma Factory, see Ref.~\cite{SeS21} for details. We can conclude, therefore, that the resonant scattering might become a powerful tool for the production of high--intensity (twisted) gamma beams with \textit{moderate values} of the TAM projection.

\section{Summary}
\label{sec:summary}

In summary, we have carried out a theoretical investigation of the resonant photon scattering by partially stripped ions. The special attention has been paid to the question of whether scattered photons are twisted if the incident radiation is twisted itself. In order to analyze this ``twistedness transfer'', we have applied the well--established S--matrix theory, which allows one to find the vector potential of scattered radiation \textit{without} projecting it onto the detector states. The expressions for this so--called \textit{evolved} state of the outgoing light have been derived for both, an incident plane wave and the twisted Bessel radiation. While the obtained expressions are general and can be used to \textit{any} resonant transition, we have illustrated their application for the simplest electric dipole $n S_{0} \to n' P_{1} \to n S_{0}$ channel. It has been shown, in particular, that the photons are scattered in the superposition of Bessel states with the TAM projections $m_f = 0, \pm 1$, if the incident radiation is twisted. Moreover, this $m_f$--distribution can be \textit{controlled} by varying the position of a ``target'' ion with respect to the incoming light beam, or by reducing the cross sectional area of ion bunches in realistic ``beam--to--beam'' collision scenarios. 

Our theoretical analysis of the resonant scattering has been performed in the ion rest frame, where the energies of the incident and outgoing photons are the same, $\omega_i = \omega_f$. However, in the laboratory frame, in which ions move with a high Lorentz factor $\gamma \gg 1$, and for the head--on collision scenario, the scattered photons will be emitted predominantly along the ion beam axis with the Doppler--boosted energy $\omega^{\rm (lab)}_f \approx 4 \gamma^2 \omega^{\rm (lab)}_i$. Along with the results, reported in the present work, it implies that the resonant photon scattering by fast--moving ions can provide an effective tool for the generation of hard x-- and even gamma--rays, carrying orbital angular momentum. Such experiments are planned, for example, at the Gamma Factory in CERN.

\section*{Acknowledgements}

The theoretical investigations, reported in Sections 2 and 3, are supported by the Government of the Russian Federation through the ITMO Fellowship and Professorship Program.
The calculations of the resonant scattering in Section 4 are supported by the Russian Science Foundation (Project No. 21-42-04412) and by the Deutsche Forschungsgemeinschaft (Project No. SU 658/5--1).

\bibliography{b21.references}{}

\end{document}